\documentclass[11pt, onecolumn]{article}

\usepackage{fancyhdr}
\usepackage[english]{babel}
\usepackage{abstract}
\usepackage[cp1251]{inputenc}

\usepackage{amsmath}  
\usepackage{amssymb} 
\usepackage{mathrsfs}

\usepackage{url}
\usepackage{euscript}
\usepackage{hyperref}

\usepackage{cyr}
\usepackage{epsfig}
\usepackage{epstopdf}

\usepackage{titlesec}
\usepackage{textcomp}

\makeatletter
\renewcommand{\@oddhead}{}
\renewcommand{\@oddfoot}{\hfill ---~\thepage~---\hfill}
\makeatother

\titleformat*{\section}{\Large\bfseries}
\titleformat*{\subsection}{\large\bfseries}

\fancypagestyle{firststyle} 
{
\fancyhead[L]{\small Published in Astronomy Letters, 2022, Vol. 48, No. 6, pp. 345-359 \\ original russian text: Pis'ma v Astronomicheskii Zhurnal, 2022, Vol. 48, No. 6, pp. 457-472.\hfill}
\fancyfoot[L]{\hfill ---~\thepage~---\hfill}
 
}

\begin{document}
\thispagestyle{firststyle}

\begin{center}
\Large Collisional Pumping of H$_2$O and CH$_3$OH Masers in C-Type Shock Waves

\vspace{0.5cm}
\large A.V. Nesterenok
\vspace{0.5cm}

\normalsize Ioffe Physical-Technical Institute, Politekhnicheskaya St. 26, Saint~Petersburg, 194021 Russia

e-mail: alex-n10@yandex.ru
\end{center}

\begin{abstract} 
\noindent
The collisional pumping of H$_2$O and CH$_3$OH masers in magnetohydrodynamic nondissociative C-type shocks is considered. A grid of C-type shock models with speeds in the range $5-70$~km~s$^{-1}$ and preshock gas densities $n_{\rm H_2,0} = 10^4-10^7$~cm$^{-3}$ is constructed. The large velocity gradient approximation is used to solve the radiative transfer equation in molecular lines. The para-H$_2$O 183.3~GHz and ortho-H$_2$O 380.1 and 448.0~GHz transitions are shown to be inverted and to have an optical depth along the shock velocity $\vert \tau \vert \sim 1$ at relatively low gas densities in the maser zone, $n_{\rm H_2} \gtrsim 10^5-10^6$~cm$^{-3}$. Higher gas densities, $n_{\rm H_2} \gtrsim 10^7$~cm$^{-3}$, are needed for efficient pumping of the remaining H$_2$O masers. Simultaneous generation of H$_2$O and class I CH$_3$OH maser emission in a shock is possible at preshock gas densities $n_{\rm H_2,0} \approx 10^5$~cm$^{-3}$ and shock speeds in the range $u_{\rm s} \approx 17.5-22.5$~km~s$^{-1}$. The possibility of detecting class I CH$_3$OH and para-H$_2$O 183.3~GHz masers in star-forming regions and near supernova remnants is investigated.
\end{abstract}

\smallskip

DOI: 10.1134/S1063773722060044

\section{Introduction}
Shocks in the interstellar medium are observed at the formation stage of stars, during their evolution, and at the final evolutionary stage of massive stars -- supernova explosions. At the star formation stage the protostellar bipolar outflows interact with the protostar envelope and the parent molecular cloud to produce shocks. After the explosion of a supernova its outer layers expand into the interstellar medium with a huge velocity, sweeping up the interstellar gas and forming a shock. In this paper we consider magnetohydrodynamic (MHD) nondissociative C-type shocks propagating in dense molecular clouds. The chemical reactions in the shock-heated gas and the sputtering of icy grain mantles change significantly the chemical composition of the gas and, at the same time, make it possible to diagnose the physical conditions through the observation of molecular and atomic lines. The more transitions of various molecules are observed in the shock-heated gas, the more accurate is the determination of physical parameters -- the shock speed and the gas density and temperature. In this paper we investigate the physical conditions under which an intense H$_2$O and CH$_3$OH maser emission arises in shocks.

Shocks can be nondissociative (C-type, there is no dissociation of molecules at the shock front) and dissociative (J-type) (Draine and McKee 1993). The shock type depends on the magnetic field strength, the gas flow speed, and the gas ionization fraction. H$_2$O maser emission can be generated in the postshock region of shocks of both types. If the gas flow speed is higher than the propagation speed of perturbations in the medium (the speed of sound and the magnetosonic speed), then a J-type shock is formed. In J-type shocks the physical parameters change in a narrow region of space with a size of the order of the mean free path of atoms and molecules. The gas in such shocks is heated to temperatures $T_{\rm g} \sim 10^5$~K, and there is a complete dissociation of molecules. Behind the shock front H$_2$ molecules are formed on dust grains, and the release of thermal energy during H$_2$ formation maintains a gas temperature of $300-400$~K. Elitzur et al. (1989) and Hollenbach et al. (2013) showed that the generation of an intense H$_2$O 22.23~GHz maser emission is possible in the warm gas behind the front of a J-type shock. However, gas temperatures $T_{\rm g} > 400$~K are needed for efficient pumping of most H$_2$O maser lines in the millimeter and submillimeter wavelength ranges (Gray et al. 2016).

If the gas flow speed is lower than the magnetosonic speed, but higher than the speed of sound for the neutral gas component, then a C-type shock is formed. In such shocks the changes in physical parameters at the shock front are determined by the diffusion of ions (and charged dust grains) and neutral gas through one another, and the gas parameters (temperature and density) undergo gradual changes. Since the kinetic energy of the gas flows is converted to thermal energy in a vast shock region, the gas is heated to temperatures much lower than those in J-type shocks: $T_{\rm g} \sim 10^3-10^4$~K. Passing through the front of a C-type shock, the gas remains molecular. In this case, the gas temperature at and behind the shock front, where the pumping of masers occurs, can be higher than that in the maser zone of J-type shocks: $T_{\rm g} \gtrsim 1000$~K (Kaufman and Neufeld 1996a). Previously it has been shown that relatively high gas densities, $n_{\rm H_2} \gtrsim 10^7$~cm$^{-3}$, are needed to pump H$_2$O masers (Neufeld and Melnick 1991; Kaufman and Neufeld 1996a; Yates et al. 1997; Gray et al. 2016, 2022). The sizes of 22.23~GHz maser emission sources also point to high gas densities in maser spots (Kaufman and Neufeld 1996a). However, Cernicharo et al. (1994, 1999) and Daniel and Cernicharo (2013) showed that some H$_2$O maser transitions (183.3, 325.1, and 380.1~GHz) could be inverted at relatively low gas densities, $n_{\rm H_2} \sim 10^5-10^6$~cm$^{-3}$.

Methanol masers are divided into two classes: class I masers with a collisional pumping mechanism and class II masers with a radiative pumping mechanism. In star-forming regions class I CH$_3$OH maser emission is generated in shocks -- the regions of interaction of the protostellar flows with the surrounding interstellar medium, in expanding HII regions (Voronkov et al. 2014). Class I CH$_3$OH masers are also observed in clouds of the central molecular zone of our Galaxy and near supernova remnants (Salii et al. 2002; Pihlstr\"{o}m et al. 2014). Methanol is formed in dark molecular clouds on dust grains in CO hydrogenation reactions (Watanabe and Kouchi 2002). At the shock front methanol falls into the gas phase as a result of the sputtering of icy grain mantles. Methanol has no formation channels in the gas phase and, therefore, methanol maser emission is generated in nondissociative C-type shocks. If the gas temperature at the shock front is sufficiently high, $T_{\rm g} \gtrsim 2000$~K, then methanol is destroyed in collisional dissociation reactions (Nesterenok 2022). The preshock gas density $n_{\rm H_2,0} \sim 10^4-10^5$~cm$^{-3}$ was shown in Nesterenok (2022) to be the most favorable condition for the emergence of an intense class I CH$_3$OH maser emission (the gas density in the maser zone is several-fold higher than the preshock gas density $n_{\rm H_2,0}$ due to gas compression in the shock). The pumping of CH$_3$OH masers can also occur at higher gas densities (McEwen et al. 2014; Leurini et al. 2016). It was shown in Nesterenok (2022) that the optical depth for CH$_3$OH transitions in a shock is small for preshock gas densities $n_{\rm H_2,0} \gtrsim 10^6$~cm$^{-3}$. At these preshock gas densities and at shock speeds when the sputtering of icy grain mantles occurs ($u_{\rm s} \gtrsim 17.5$~km~s$^{-1}$), the gas temperature at the shock front is $T_{\rm g} \gtrsim 2000$~K and there is a (partial) dissociation of methanol molecules. Thus, class I CH$_3$OH masers and some H$_2$O transitions have a pumping regime at relatively low gas densities.

This paper is a continuation of our study of maser pumping in shocks begun in Nesterenok (2020, 2021, 2022). The collisional pumping of OH masers at 1720~MHz in shocks near supernova remnants was considered in Nesterenok (2020). In Nesterenok (2021, 2022) we considered the pumping of class I CH$_3$OH masers and studied the coexistence of CH$_3$OH and OH masers in the same source. In this paper we investigate the collisional pumping of H$_2$O masers in C-type shocks for preshock gas densities $n_{\rm H_2,0} = 10^4-10^7$~cm$^{-3}$, and consider the coexistence of H$_2$O and class I CH$_3$OH masers in the same
source.

\section{C-type shock model}
The model of a steady-state C-type shock propagating in a dense molecular cloud was developed in Nesterenok (2018) and Nesterenok et al. (2019). The numerical simulations consist of two parts: (1) the simulations of the chemical evolution of the dark molecular cloud and (2) the simulations of the shock propagation. At the start of the simulations of the cloud's chemical evolution the H atoms are assumed to be bound into H$_2$ molecules, while all of the remaining elements are in the atomic or ionized state (Nesterenok 2022). A detailed description of all the chemical processes that are taken into account in the numerical simulations and a description of the dynamics of the gas components in the shock (neutral gas, ions, electrons, and dust grains) are given in Nesterenok (2018). The chemical reactions that determine the methanol concentration are discussed in Nesterenok (2022).

As a starting point for the shock simulations we chose the age of the molecular cloud $t_0$ at which the methanol abundance relative to the hydrogen nuclei in the icy mantles of dust grains is 10$^{-5}$. This age depends on the gas density and the cosmic-ray ionization rate. At the same time, the relative abundance of H$_2$O molecules at $t_0$ slightly differs for different gas densities and gas ionization rates and is $5 \times 10^{-5}-10^{-4}$. According to the observational data, the relative abundance of H$_2$O and CH$_3$OH molecules in the icy mantles of dust grains in molecular clouds lies in the range $(1-8) \times 10^{-5}$ and $(0-1.5) \times 10^{-5}$, respectively (Boogert et al. 2015). The relative CH$_3$OH abundance adopted in our calculations corresponds to the upper limit of the observed values.

To estimate the preshock magnetic field, we used a power-law dependence of the magnetic field on gas density (Dudorov 1991; Crutcher et al. 2010):

\begin{equation}
B = \beta B_0 \left( n_\mathrm{H,tot}/n_0 \right)^{\alpha},
\label{eq_magn_field}
\end{equation}

\noindent
where the values of the parameters are as follows: $n_0 = 300$~cm$^{-3}$, $B_0 = 10$~$\mu$G, $\alpha = 0.65$, and the number density of hydrogen nuclei $n_{\rm H,tot} \geq n_0$. According to the Zeeman molecular line splitting observations, the magnetic field in molecular clouds varies in a wide range, $0 < \beta \leq 1$ (Crutcher et al. 2010). In most of our calculations we use $\beta = 1$ and the direction of the magnetic field is perpendicular to the shock velocity. The results of our shock model computations in which $\beta = 0.5$ are also presented.

In cold molecular clouds the ortho-H$_2$-to-para-H$_2$ conversion time scale can be larger than the cloud evolution time, and the ortho-/para-H$_2$ ratio has no time to reach its equilibrium value. The initial ortho-/para-H$_2$ ratio was chosen to be 0.1 -- some arbitrary low value. In our calculations we take into account the following processes through which the para-/ortho-H$_2$ interconversion occurs: H$_2$--H collisions, H$_2$--H$^+$ collisions, and H$_2$ formation on dust grains (Nesterenok et al. 2019). When simulating the chemical evolution of a cold molecular cloud, the ortho-/para-H$_2$ ratio changes slowly toward its equilibrium value. In this case, collisions with H$^+$ are the main para-/ortho-H$_2$ interconversion channel.

The cosmic-ray ionization rate in most of our calculations was set equal to $\zeta_{\rm H_2} = 3 \times 10^{-17}$~s$^{-1}$, corresponding to the gas ionization rate in cold molecular clouds away from the sources of ionizing radiation (Dalgarno 2006). We also present the results of our shock model computations in which the gas ionization rate was set equal to $\zeta_{\rm H_2} = 3 \times 10^{-15}$~s$^{-1}$. This value may be considered as a typical gas ionization rate in molecular clouds in the vicinity of supernova remnants and in clouds in the central molecular zone of our Galaxy (Shingledecker et al. 2016). The shock speeds varied from 5~km~s$^{-1}$ to the limiting C-type shock speed. The limiting speed is determined from the condition of almost complete H$_2$ dissociation. The limiting speeds are approximately 70, 45, 30, and 30~km~s$^{-1}$ for the preshock gas densities $n_{\rm H_2,0} = 10^4$, $10^5$, $10^6$, and $10^7$~cm$^{-3}$, respectively (for the gas ionization rate $\zeta_{\rm H_2}= 3 \times 10^{-17}$~s$^{-1}$). Table~1 gives the parameters that were used in our numerical simulations of shocks.

~\\
~\\
\begin{tabular}{l@{\quad\quad}@{\quad\quad}l}
\multicolumn{2}{l}{\large\bf Table 1. Shock parameters} \\ [5pt]
\hline \\ [-2ex]
Preshock gas density, $n_{\rm H_2,0}$ & $10^4 - 10^7$~cm$^{-3}$ \\ [5pt]
Shock speed, $u_{\rm s}$ & $5-70$~km~s$^{-1}$ \\ [5pt]
Cosmic-ray ionization rate, $\zeta_{\rm H_2}$ & $3 \times 10^{-17}$, $3 \times 10^{-15}$~s$^{-1}$ \\ [5pt]
Initial ortho-H$_2$/para-H$_2$ ratio & 0.1 \\ [5pt]
Parameter characterizing the magnetic field strength, $\beta$ & 0.5, 1 \\ [5pt]
Turbulent velocity, $v_{\rm turb}$ & 0.3~km~s$^{-1}$ \\ [5pt]
Initial relative CH$_3$OH abundance in the icy mantles of dust grains & $10^{-5}$ \\ [5pt]
Initial relative H$_2$O abundance in the icy mantles of dust grains & $(5-10) \times 10^{-5}$ \\ [5pt]

\hline
\end{tabular}
~\\
~\\
The parameter $\beta$ is defined in Eq. (\ref{eq_magn_field}).
~\\

\section{Calculation of molecular energy level populations}
\subsection{Collisional Rate Coefficients and Spectroscopic Data}

The energies of rotational levels and the Einstein coefficients for the H$_2$O molecule were taken from the HITRAN 2020 database (Gordon et al. 2022). In our calculations we took into account 150 rotational energy levels of the para-H$_2$O molecule and 150 energy levels of the ortho-H$_2$O molecule belonging to the ground and first excited vibrational states of the molecule. The energy of the highest H$_2$O level considered is 4500~K. The collisional rate coefficients for transitions between H$_2$O energy levels in collisions of H$_2$O with H$_2$ and electrons were calculated in Faure et al. (2007) and Faure and Josselin (2008). Four data sets for collisions between ortho-/para-H$_2$O and ortho-/para-H$_2$ for the gas temperature range $20-2000$~K are given in Faure et al. (2007), with the transitions between 45 lower energy levels of each H$_2$O spin isomer being considered. For the remaining transitions in our calculations we used data from Faure and Josselin (2008). In the calculations of these collisional rate coefficients the ortho-/para-H$_2$ ratio was initially set equal to 3. The collisional rate coefficients for transitions between H$_2$O levels in collisions of H$_2$O with He atoms were taken from Green et al. (1993) and Nesterenok (2013). The collisional rate coefficients for H$_2$O transitions in collisions with H atoms for 45 lower rotational H$_2$O levels were taken from Daniel et al. (2015). The collisions of H$_2$O with H atoms become significant when the shock speed is close to the limiting C-type shock speed and there is a partial dissociation of H$_2$ molecules at the shock front. The abundance of electrons relative to the hydrogen nuclei in the molecular gas is $x_e \sim 10^{-8}-10^{-7}$ for $\zeta_{\rm H_2} \sim 10^{-16}$~s$^{-1}$ and $n_{\rm H_2} \sim 10^4-10^5$~cm$^{-3}$, and decreases with increasing gas density. The collisions of H$_2$O with electrons are insignificant.

The spectroscopic data and the data on the collisional rate coefficients that were used in our calculations of the CH$_3$OH energy level populations are described in Nesterenok (2016). In our calculations we do not use the extrapolation of the collisional rate coefficients for high gas temperatures -- the rate coefficients are assumed to be constant at temperatures above the maximum temperature for which data are available. The sensitivity of the results of our calculations of the CH$_3$OH energy level populations to the collisional rate coefficients at high temperatures was analyzed in Nesterenok (2022). The spin-isomer abundance ratio was assumed to be the following: ortho-/para-H$_2$O = 3 (Emprechtinger et al. 2013) and A-/E-CH$_3$OH = 1 (Nesterenok 2022).

\subsection{Basic Formulas}
In this paper we use the same method of calculating the molecular level populations in a shock as that in Nesterenok (2020, 2022). Below, we briefly outline the ideas of the method. The shock profile obtained as a result of our numerical simulations is divided into layers. For each layer we calculate the H$_2$O and CH$_3$OH energy level populations. The system of equations for the energy level populations of a molecule at some distance $z$ in the shock is

\begin{equation}
\begin{array}{c}
\displaystyle
\sum_{k=1, \, k \ne i}^M \left( R_{ki} + C_{ki} \right) n_k(z) - n_i(z) \sum_{k=1, \, k \ne i}^M \left( R_{ik} + C_{ik} \right)=0, \quad i=1,...,M-1, \\
\displaystyle
\sum_{i=1}^M n_i(z)=1,
\end{array}
\label{eq1}
\end{equation}

\noindent
Here, $M$ is the total number of energy levels, $R_{ik}$ is the probability of radiative transitions from level $i$ to level $k$, and $C_{ik}$ is the probability of collisional transitions. The probabilities of radiative transitions are as follows:

\begin{equation}
\begin{array}{c}
\displaystyle
R_{ik}^{\downarrow}=B_{ik}J_{ik}+A_{ik}, \quad i > k, \\[10pt]
\displaystyle
R_{ik}^{\uparrow}=B_{ik}J_{ik}, \quad i < k,
\end{array}
\label{eq_rad_prob}
\end{equation}

\noindent
where $A_{ik}$ and $B_{ik}$ are the Einstein coefficients for spontaneous and stimulated emission, respectively, and $J_{ik}$ is the radiation intensity averaged over the direction and the line profile. To calculate the radiation intensity, we used the large velocity gradient method (Hummer and Rybicki 1985). This method gives a good approximation if the length scale of the change in physical parameters is much greater than the Sobolev length:

\begin{equation}
\Delta z_\mathrm{S} = u_\mathrm{D} \left\vert \frac{\mathrm{d}u(z)}{\mathrm{d}z} \right\vert^{-1}
\end{equation}

\noindent
where $u(z)$ is the gas velocity and $u_D$ is the line profile width in velocity units (Nesterenok 2020, 2022). The overlap between CH$_3$OH and H$_2$O lines was ignored since for the masers under consideration the line overlap has a minor effect on pumping (McEwen et al. 2014; Gray et al. 2016). The absorption of radiation in molecular lines by dust was taken into account (Hummer and Rybicki 1985; Nesterenok 2016). The dust temperature behind the shock front, where the pumping of masers occurs, is much lower than the gas temperature, $T_{\rm d} << T_{\rm g}$. The maximum dust temperature is reached at the shock peak and is 65~K for the model with parameters $n_{\rm H_2,0} = 10^7$~cm$^{-3}$ and $u_{\rm s} = 30$~km~s$^{-1}$. The dust radiation was ignored in our calculations of the radiation intensity in molecular lines. The system of equations (\ref{eq1}) was solved by the iteration method.

Once the system of equations (\ref{eq1}) for the molecular energy level populations had been solved, we calculated the gain for transitions with level population inversion. The expression for the gain (which is equal to the absorption coefficient with the opposite sign) for a $i \to k$ transition in the case of a plane-parallel gas-dust cloud is

\begin{equation}
\displaystyle
\gamma_{ik}\left( z, \mu, \nu \right)=\frac{\lambda^2}{8 \pi} A_{ik} n_{\rm m}(z) \, \left(n_i(z) - \frac{g_i}{g_k}n_k(z) \right) \phi \left( z, \mu, \nu \right) - \kappa_{\rm c}(z),
\end{equation}

\noindent
where $\mu$ is the cosine of the angle between the gas flow direction in the shock and the line of sight, $n_{\rm m}(z)$ is the number density of molecules (ortho- or para-H$_2$O, A- or E-type CH$_3$OH spin isomers) at distance $z$ in the shock, $g_i$ and $g_k$ are the statistical weights of the energy levels, and $\kappa_{\rm c}(z)$ is the dust absorption coefficient. The spectral profile of the emission and absorption coefficients in the laboratory frame of reference is given by the expression

\begin{equation}
\phi(z,\mu, \nu) = \tilde{\phi}_\mathrm{ik} \left[ \nu - \nu_\mathrm{ik} \mu u(z) / c \right]
\label{eq_profile_lab}
\end{equation}

\noindent
where $\nu_{ik}$ is the transition frequency and $\tilde{\phi}_\mathrm{ik}(\nu)$ is the normalized spectral profile in the frame of reference associated with the gas flow. For the ortho-H$_2$O $6_{16} \to 5_{23}$ transition at 22.23~GHz it is necessary to take into account the additional line profile broadening due to the hyperfine splitting of energy levels (Varshalovich et al. 2006; Nesterenok and Varshalovich 2011). The spectral profile of the emission and absorption coefficients in this line is the sum of six components with different intensities. At a gas temperature $T_{\rm g} \gtrsim 150$~K the components merge into a single asymmetric profile. For ortho-H$_2$O transitions in the millimeter and submillimeter wavelength ranges the splitting is small compared to the Doppler line profile width.

The optical depth in the line for which there is level population inversion is

\begin{equation}
\displaystyle
\vert \tau_{\mu}(\nu) \vert = \frac{1}{\mu}\int \, \mathrm{d}z \, \gamma_\mathrm{ik}(z,\mu, \nu).
\label{eq_tau}
\end{equation}

\noindent
The parameter $a = 1/\mu$ is equal to the ratio of the amplification path length along the line of sight to the shock width. If the shock is seen edge-on, then $a$ is large and the maser emission is most intense. In theoretical works $a \sim 10$ is assumed to explain the emission of bright H$_2$O masers (see, e.g., Kaufman and Neufeld 1996a). The optical depth increases with $a$ faster than $\propto a$ due to the dependence of the spectral line profile on $\mu$, see Eq. (\ref{eq_profile_lab}) (Nesterenok 2021). The maximum of (\ref{eq_tau}) at fixed $\mu$ is reached at the line center.

For the brightness temperature in a maser line one
can write

\begin{equation}
T_{\rm b} = T_{\rm bg} \exp\left(\vert \tau \vert \right),
\label{eq_br_temp}
\end{equation}

\noindent
where $T_{\rm bg}$ is the background radiation temperature and $\vert \tau \vert$ is the absolute value of the optical depth in the maser line. As the radiation intensity in the maser line increases, the rate of induced transitions (the term proportional to $J_{ik}$ in Eqs. (\ref{eq_rad_prob})) becomes comparable to the rates of collisional and radiative transitions to other levels. In this case, the maser passes to the regime of saturation, and the exponential amplification law changes to a linear one (Strelnitskii 1975). In our paper we did not consider the maser amplification in the saturated regime.

\section{Results}
\subsection{Physical Conditions in the Maser Formation Region}

\begin{figure}[h]
\centering
\includegraphics[width = 0.8\textwidth]{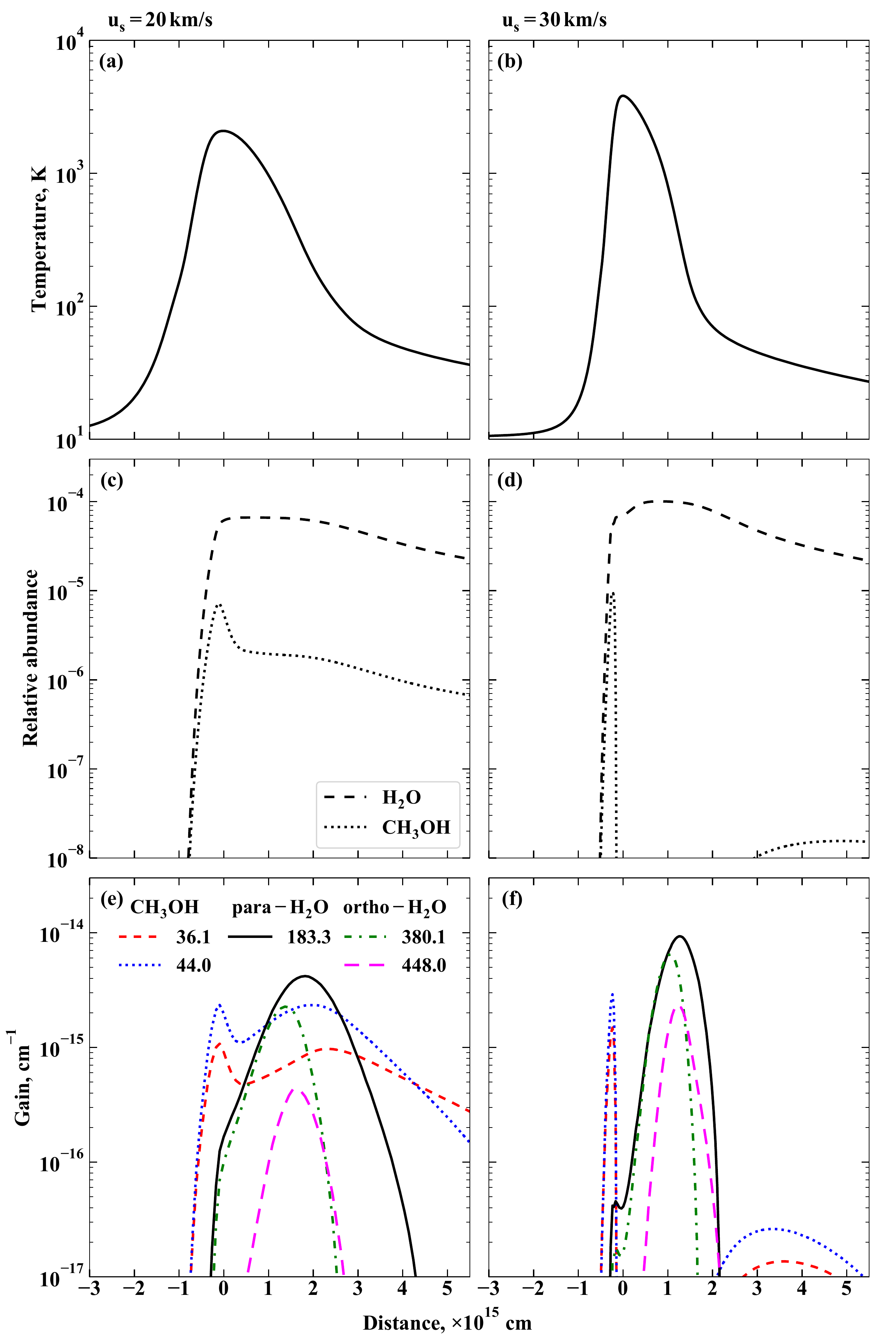}
\caption{(a, b) Gas temperature in the shock versus distance; (c, d) the abundance of H$_2$O and CH$_3$OH molecules in the gas relative to the abundance of hydrogen nuclei; (e, f) the gain for H$_2$O and CH$_3$OH maser transitions. The results of our calculations for the preshock gas density $n_{\rm H_2,0} = 10^5$~cm$^{-3}$, the cosmic-ray ionization rate $\zeta_{\rm H_2} = 3 \times 10^{-17}$~s$^{-1}$, and two shock speeds, 20 (left) and 30 (right)~km~s$^{-1}$, are shown. The zero-point on the horizontal axis is the point where the gas temperature attains maximum.}
\end{figure}

Figure~1 shows plots of the temperature, the relative abundance of H$_2$O and CH$_3$OH molecules, and the gain in maser lines as functions of the distance along the gas flow in the shock. The results are presented for two shock models with speeds $u_{\rm s} = 20$ and 30~km~s$^{-1}$. The preshock gas density for both shock models is $n_{\rm H_2,0} = 10^5$~cm$^{-3}$. The gas temperature rises rapidly to its maximum value (2100 and 3800~K for $u_{\rm s} = 20$ and 30~km~s$^{-1}$, respectively). Subsequently, the gas temperature falls slowly as a result of the reduction in the gas heating rate and of the gas cooling through radiation in molecular lines. At the shock front the icy mantles of dust grains are sputtered, and a sharp rise in the relative abundance of H$_2$O and CH$_3$OH in the gas phase is observed. Behind the shock front the relative abundance of molecules in the gas falls slowly as a result of adsorption on dust grains. The time scale of this process is $t \sim 10^3$~yr for $n_{\rm H_2} = 10^6$~cm$^{-3}$ and $T_{\rm g} = 50$~K. In the hot gas at the shock front there is destruction of methanol molecules in reactions with H atoms and collisional dissociation reactions. For the shock speed $u_{\rm s} = 20$~km~s$^{-1}$ methanol is destroyed incompletely and the CH$_3$OH-to-H$_2$O ratio in the postshock region is 0.03. At a higher speed, $u_{\rm s} = 30$~km~s$^{-1}$, the methanol molecules are destroyed completely in the hot gas at the shock front. On the other hand, the higher the shock speed, the higher the relative H$_2$O abundance in the cooling gas behind the shock front: the H$_2$O column density from the shock peak to the region where the gas temperature drops below 30~K is $N_{\rm H_2O} = 3 \times 10^{17}$ and $5 \times 10^{17}$~cm$^{-2}$ for $u_{\rm s} = 20$ and 30~km~s$^{-1}$, respectively. This is because O and OH forming in reactions of molecules with H atoms and collisional dissociation reactions (the destruction of CH$_3$OH, CO$_2$, and other molecules) turn into H$_2$O.

In Figs. 1e and 1f the gain is plotted against the distance for the methanol E~$4_{-1} \to 3_0$ 36.1~GHz and A$^+$~$7_0 \to 6_1$ 44.0~GHz lines, the para-H$_2$O $3_{13} \to 2_{20}$ 183.3~GHz line, and the ortho-H$_2$O $4_{14} \to 3_{21}$ 380.1~GHz and $4_{23} \to 3_{30}$ 448.0~GHz lines. The size of the shock region where the gain in the para-H$_2$O line at 183.3~GHz drops by a factor of 2 from its maximum value is $1.5 \times 10^{15}$~cm for the shock speed $u_{\rm s} = 20$~km~s$^{-1}$ and half as much for $u_{\rm s} = 30$~km~s$^{-1}$. Within this region the gas temperature drops by a factor of 10 (from $T_{\rm g} \approx 1000$~K to 100~K), whereas the gas density increases by a factor of 2 (from $3 n_{\rm H_2,0}$ to $6n_{\rm H_2,0}$ for $u_{\rm s} = 20$~km~s$^{-1}$) and the absolute value of the gas velocity gradient decreases by a factor of 5. The gas temperature is the parameter with the largest gradient and, therefore, the position of the gain peak for the maser transitions is determined mainly by the change in gas temperature. In the region where the gain in the 183.3~GHz line reaches its maximum, the gas temperature is $250-300$~K and the gas density is $n_{\rm H_2} \approx 5n_{\rm H_2,0}$. Higher temperatures are needed for efficient pumping of the 380.1~GHz transition, and the 380.1~GHz maser emission region must be more compact. In the region where the gain in the 380.1~GHz line reaches its maximum the gas temperature is $\approx 600$~K. The population inversion in the 380.1~GHz line vanishes as soon as the gas temperature drops below 100~K. For the shock speed $u_{\rm s} = 20$~km~s$^{-1}$ there is energy level population inversion for the CH$_3$OH 36.1 and 44.0~GHz transitions in the wide region from the shock front to the far postshock zone, where the gas temperature drops to 30~K. 

The dust absorption coefficient at 183.3~GHz is $\kappa_{\rm c} \approx (5-10) \times 10^{-21}$~cm$^{-1}$ in the shock region, where the gain in the 183.3~GHz maser line is at a maximum (for $n_{\rm H_2,0} = 10^5$~cm$^{-3}$). Thus, the dust absorption for the maser transitions is negligible.

\subsection{Optical Depth in H$_2$O and CH$_3$OH Maser Lines}

\begin{figure}[h]
\centering
\includegraphics[width = 0.85\textwidth]{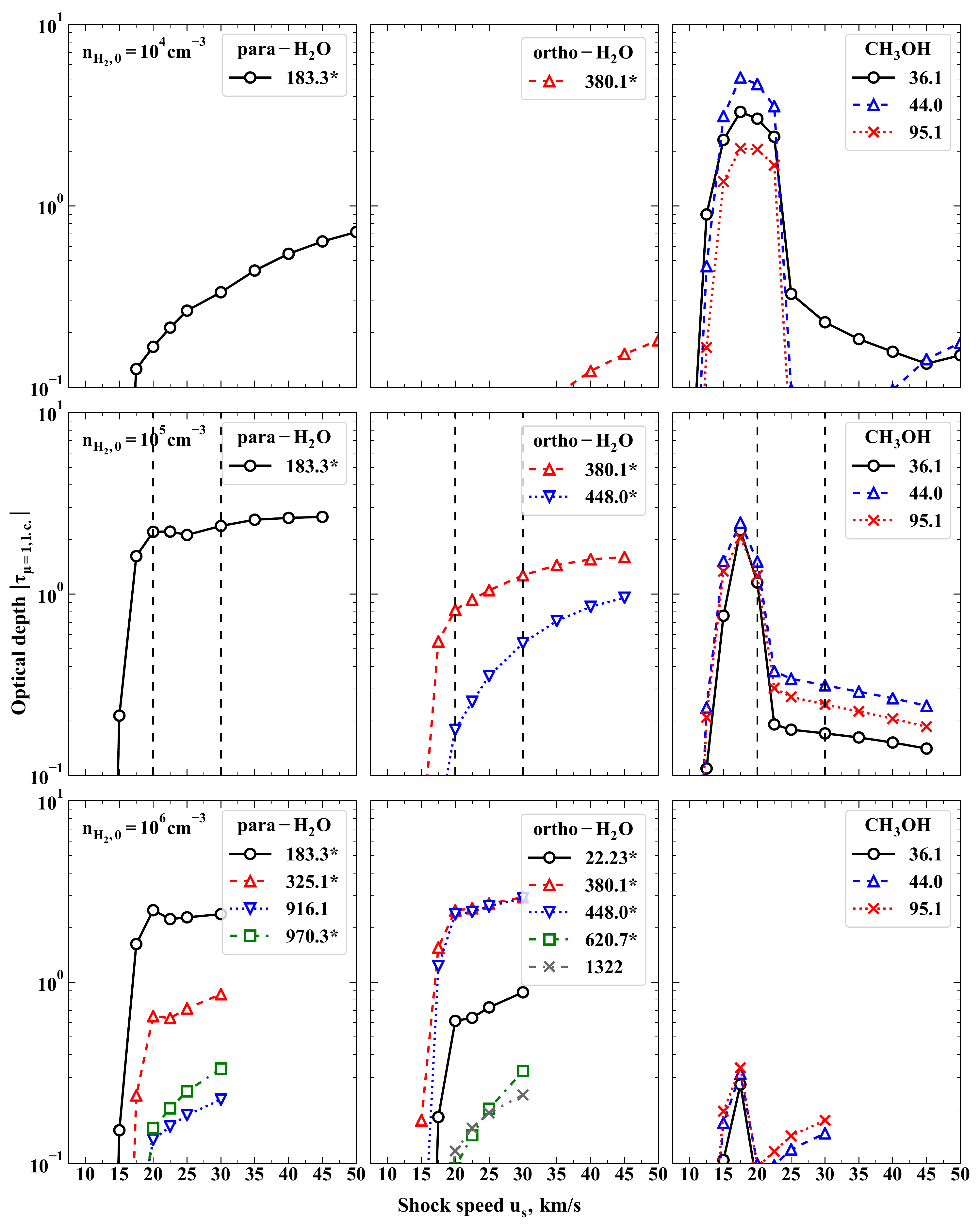}
\caption{Optical depth (absolute value) at the line center along the gas flow $\vert \tau_{\rm \mu=1, l.c.}\vert$ for the H$_2$O and CH$_3$OH maser transitions versus shock speed $u_{\rm s}$. Each point corresponds to one shock model, with the shock speed being along the horizontal axis. The calculations are presented for three preshock gas densities: $n_{\rm H_2,0} = 10^4$ (upper row), $10^5$ (middle row), and $10^6$ (lower row) cm$^{-3}$. The left, middle, and right columns show the para-H$_2$O, ortho-H$_2$O, and CH$_3$OH transitions, respectively. The H$_2$O transitions whose emission was observed in astrophysical objects are designated by an asterisk (Neufeld et al. 2017; Pereira-Santaella et al. 2017). The results of our calculations for the CH$_3$OH E~$4_{-1} \to 3_0$ 36.1~GHz, A$^+$ $7_0 \to 6_1$ 44.0~GHz, and A$^+$ $8_0 \to 7_1$ 95.1~GHz lines are shown. The vertical lines in the middle row designate the shock models for which the results of our calculations are presented in Fig.~1.}
\end{figure}

Figure~2 shows the results of our calculations of the optical depth at the line center along the gas flow $\vert \tau_{\rm \mu=1, l.c.}\vert$ for the H$_2$O and CH$_3$OH maser transitions. The calculations are presented for the preshock gas densities $n_{\rm H_2,0} = 10^4$, $10^5$, and $10^6$~cm$^{-3}$. This figure shows all of the inverted H$_2$O transitions for which, according to our calculations, the optical depth $\vert \tau_{\rm \mu=1, l.c.}\vert > 0.1$. According to our calculations, an optimal condition for the pumping of methanol masers is the gas density range $n_{\rm H_2,0} = 10^4 - 10^5$~cm$^{-3}$. At such gas densities the optical depth $\vert \tau_{\rm \mu=1, l.c.}\vert \sim 1$ for the para-H$_2$O transition at 183.3~GHz and the ortho-H$_2$O transitions at 380.1 and 448.0~GHz. At gas densities $n_{\rm H_2,0} \geq 10^6$~cm$^{-3}$ the optical depth for the CH$_3$OH maser transitions is small due to the destruction of methanol molecules in the hot dense gas at the shock front (Nesterenok 2022). At the same time, for many H$_2$O transitions the optical depth $\vert \tau_{\mu=1, l.c.}\vert \geq 0.1$ at such a gas density.

Figure 3 shows the results of our calculations of the optical depth for the H$_2$O maser transitions for the preshock gas density $n_{\rm H_2,0} = 10^7$~cm$^{-3}$. In this case, there is energy level population inversion for a much larger number of H$_2$O transitions than in the case of lower gas densities. The list of transitions is given in Table~2; all transitions belong to the ground vibrational H$_2$O state. According to our calculations, the para-H$_2$O 183.3, 325.1, and 970.3~GHz transitions and the ortho-H$_2$O 22.23, 380.1, 448.0, 620.7, 1296, and 1322~GHz transitions have an optical depth $\vert \tau_{\rm \mu=1, l.c.}\vert > 1$ at least in one of the shock models. If the shock is seen edge-on ($a \sim 10$), then these transitions are strong masers. The well-known ortho-H$_2$O 321.2, and 439.1~GHz maser transitions have an optical depth $\vert \tau_{\rm \mu=1, l.c.}\vert \sim 0.1$ (Fig. 3). In our calculations we obtained a small optical depth $\vert \tau_{\rm \mu=1, l.c.}\vert < 0.1$ (or the absence of energy level population inversion) for the maser transitions of excited vibrational H$_2$O states.

In Fig.~4 the optical depth $\vert \tau_{\rm \mu=1, l.c.}\vert$ for the H$_2$O and CH$_3$OH maser transitions is plotted against the preshock gas density $n_{\rm H_2,0}$, the shock speed in all calculations was set equal to $u_{\rm s} = 20$~km~s$^{-1}$. As the preshock gas density increases, the optical depth in the CH$_3$OH lines decreases, while for the H$_2$O transitions it increases. At gas densities $n_{\rm H_2,0} \approx 10^5$~cm$^{-3}$ a coexistence of class I CH$_3$OH masers and H$_2$O 183.3 and 380.1~GHz masers is possible. The higher the preshock gas density, the narrower the shock front. The length of the postshock region at which the gain in the 183.3~GHz line drops by a factor of 2 from its maximum value is $\approx 10^{14}$~cm for $n_{\rm H_2,0} = 10^7$~cm$^{-3}$ -- an order of magnitude less than that for a preshock gas density of $10^5$~cm$^{-3}$.

\begin{figure}[h]
\centering
\includegraphics[width = 0.9\textwidth]{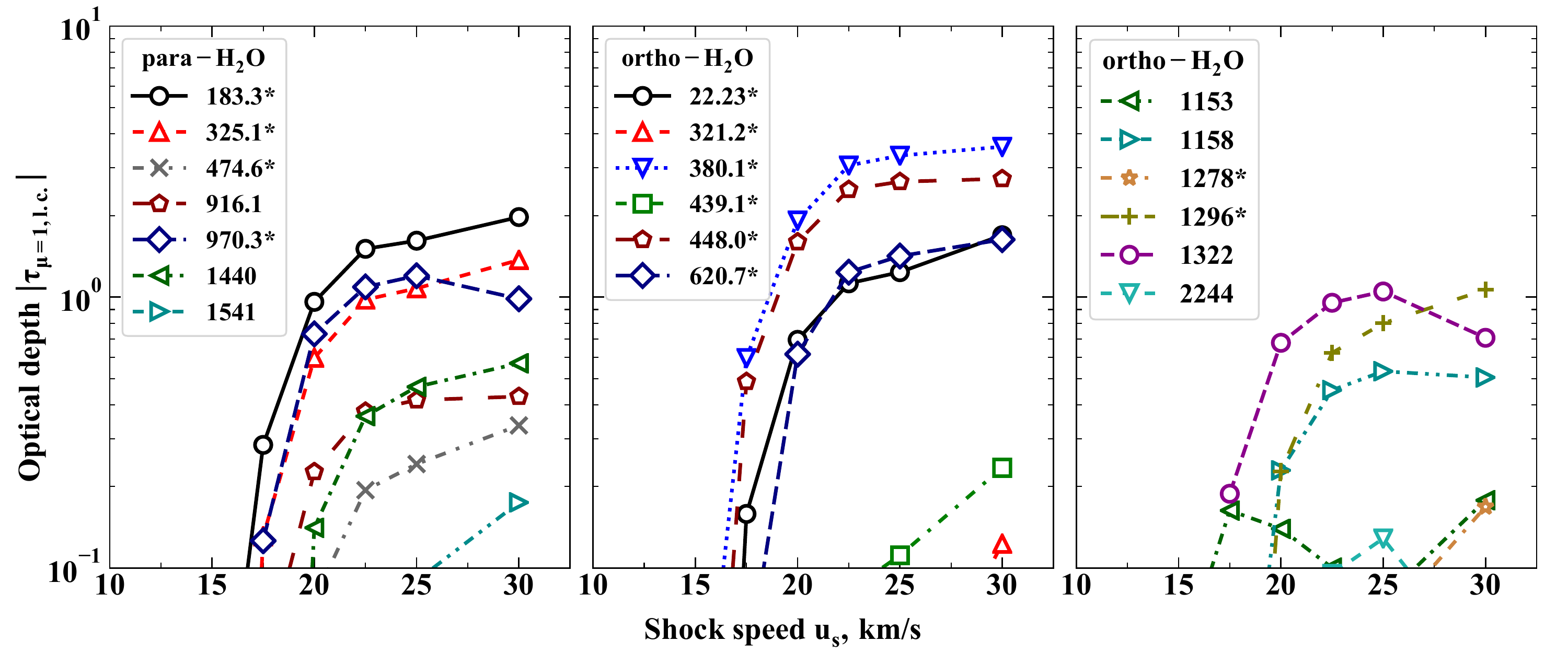}
\caption{Optical depth at the line center along the gas flow $\vert \tau_{\rm \mu=1, l.c.}\vert$ for the H$_2$O maser transitions versus shock speed. The preshock gas density is $n_{\rm H_2,0} = 10^7$~cm$^{-3}$. The left panel shows the results for the para-H$_2$O lines; the remaining two panels show the results for the ortho-H$_2$O lines.}
\end{figure}

\begin{figure}[h]
\centering
\includegraphics[width = 0.6\textwidth]{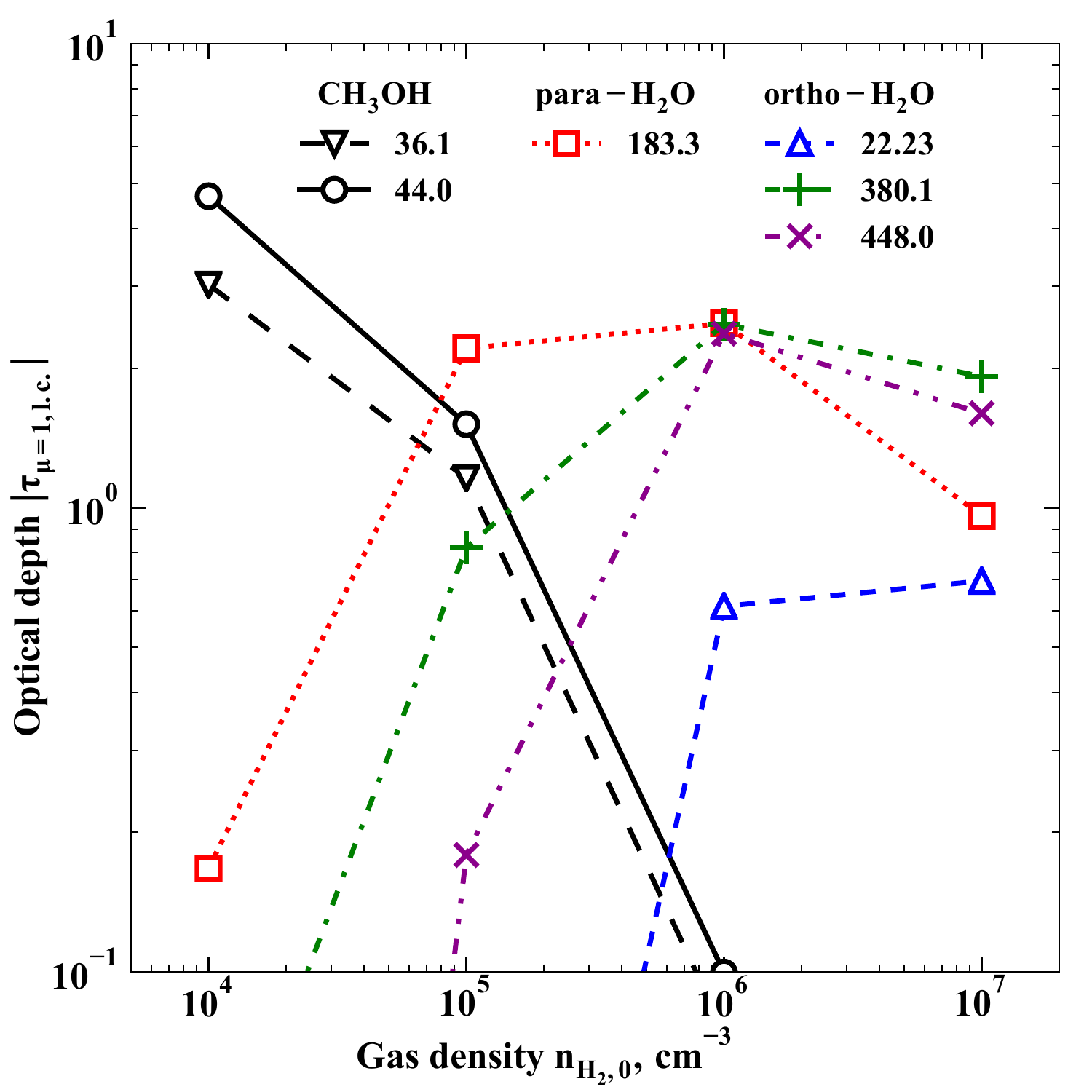}
\caption{Optical depth at the line center along the gas flow $\vert \tau_{\rm \mu=1, l.c.}\vert$ for the H$_2$O and CH$_3$OH maser transitions versus preshock gas density $n_{\rm H_2,0}$. The shock speed in all calculations was set equal to $u_{\rm s} = 20$~km~s$^{-1}$.}
\end{figure}

~\\
~\\
\begin{tabular}{llll}
\multicolumn{2}{l}{\large\bf Table 2. H$_2$O transitions.} \\ [5pt]
\hline
\multicolumn{2}{c}{ortho-H$_2$O} & \multicolumn{2}{c}{para-H$_2$O} \\ [3pt]
\hline
$6_{1\,6} \to 5_{2\,3}$ & 22.23* & $3_{1\,3} \to 2_{2\,0}$ & 183.3* \\ [3pt]
$10_{2\,9} \to 9_{3\,6}$ & 321.2* & $5_{1\,5} \to 4_{2\,2}$ & 325.1* \\ [3pt]
$4_{1\,4} \to 3_{2\,1}$ & 380.1* &  $5_{3\,3} \to 4_{4\,0}$ & 474.6* \\ [3pt]
$6_{4\,3} \to 5_{5\,0}$ & 439.1* & $4_{2\,2} \to 3_{3\,1}$ & 916.1 \\ [3pt]
$4_{2\,3} \to 3_{3\,0}$ & 448.0* & $5_{2\,4} \to 4_{3\,1}$ & 970.3* \\ [3pt]
$5_{3\,2} \to 4_{4\,1}$ & 620.7* & $7_{2\,6} \to 6_{3\,3}$ & 1440 \\ [3pt]
$3_{1\,2} \to 2_{2\,1}$ & 1153 & $6_{3\,3} \to 5_{4\,2}$ & 1541 \\ [3pt]
$6_{3\,4} \to 5_{4\,1}$ & 1158 &  &  \\ [3pt]
$7_{4\,3} \to 6_{5\,2}$ & 1278* &  &  \\ [3pt]
$8_{2\,7} \to 7_{3\,4}$ & 1296* &  &  \\ [3pt]
$6_{2\,5} \to 5_{3\,2}$ & 1322 &  &  \\ [3pt]
$8_{3\,6} \to 7_{4\,3}$ & 2244 &  &  \\ [3pt]
\hline
\end{tabular}
~\\
~\\
The list of transitions for which the optical depth $\vert \tau_{\rm \mu=1, l.c.}\vert > 0.1$ at least in one of the shock models for the preshock gas density $n_{\rm H_2,0} = 10^7$~cm$^{-3}$. All transitions belong to the ground vibrational H$_2$O state. The H$_2$O transitions whose emission was observed in astrophysical objects are designated by an asterisk (Neufeld et al. 2017; Pereira-Santaella et al. 2017). The frequencies are given in GHz, "truncated" values are used.
~\\

\subsection{Effect of the ortho-/para-H$_2$ Ratio on Maser Pumping}
In the hot gas at the shock front H$_2$--H collisions are the main para-H$_2$/ortho-H$_2$ interconversion mechanism. If the gas temperature and the number density of H atoms in the gas are sufficiently high, then the ortho-/para-H$_2$ ratio has time to reach its equilibrium value determined by the gas temperature (Nesterenok 2019). For shock speeds less than some value of u$_0$ the para-H$_2$-to-ortho-H$_2$ conversion in the heated gas at the shock front is inefficient. In this case, the main collisional partner of molecules in collisions is para-H$_2$. The value of $u_0$ is $\approx 30$ and 20~km~s$^{-1}$ for the preshock gas densities $n_{\rm H_2,0} = 10^4$ and $10^7$~cm$^{-3}$, respectively ($\zeta_{\rm H_2} =  3 \times 10^{-17}$~s$^{-1}$). In particular, in a shock with parameters $n_{\rm H_2,0} = 10^5$~cm$^{-3}$ and $u_{\rm s} = 22.5$~km~s$^{-1}$ the ortho-/para-H$_2$ ratio increases from 0.02 to 0.5 as the gas passes through the shock front. We performed calculations in which the ortho-/para-H$_2$ ratio was initially set equal to 3 in the shock model with these parameters. The difference of the results of our calculations for the optical depth in the para-H$_2$O 183.3~GHz line is about 3\%. This result is explained by the fact that the H$_2$O--H$_2$ collisional rate coefficients have a small difference for ortho- and para-H$_2$ for gas temperatures above 300~K (Faure et al. 2007). For the CH$_3$OH 36.1, 44.0, and 95.1~GHz transitions the optical depth is smaller by $\approx 30$\% for an ortho-/para-H$_2$ ratio of 3. The influence of the ortho-/para-H$_2$ ratio on the pumping of H$_2$O masers is minor and significant for CH$_3$OH masers.

\subsection{Effect of the Gas Ionization Rate and the Magnetic Field Strength on the Generation of H$_2$O and CH$_3$OH Maser Emission}

\begin{figure}[h]
\centering
\includegraphics[width = 0.9\textwidth]{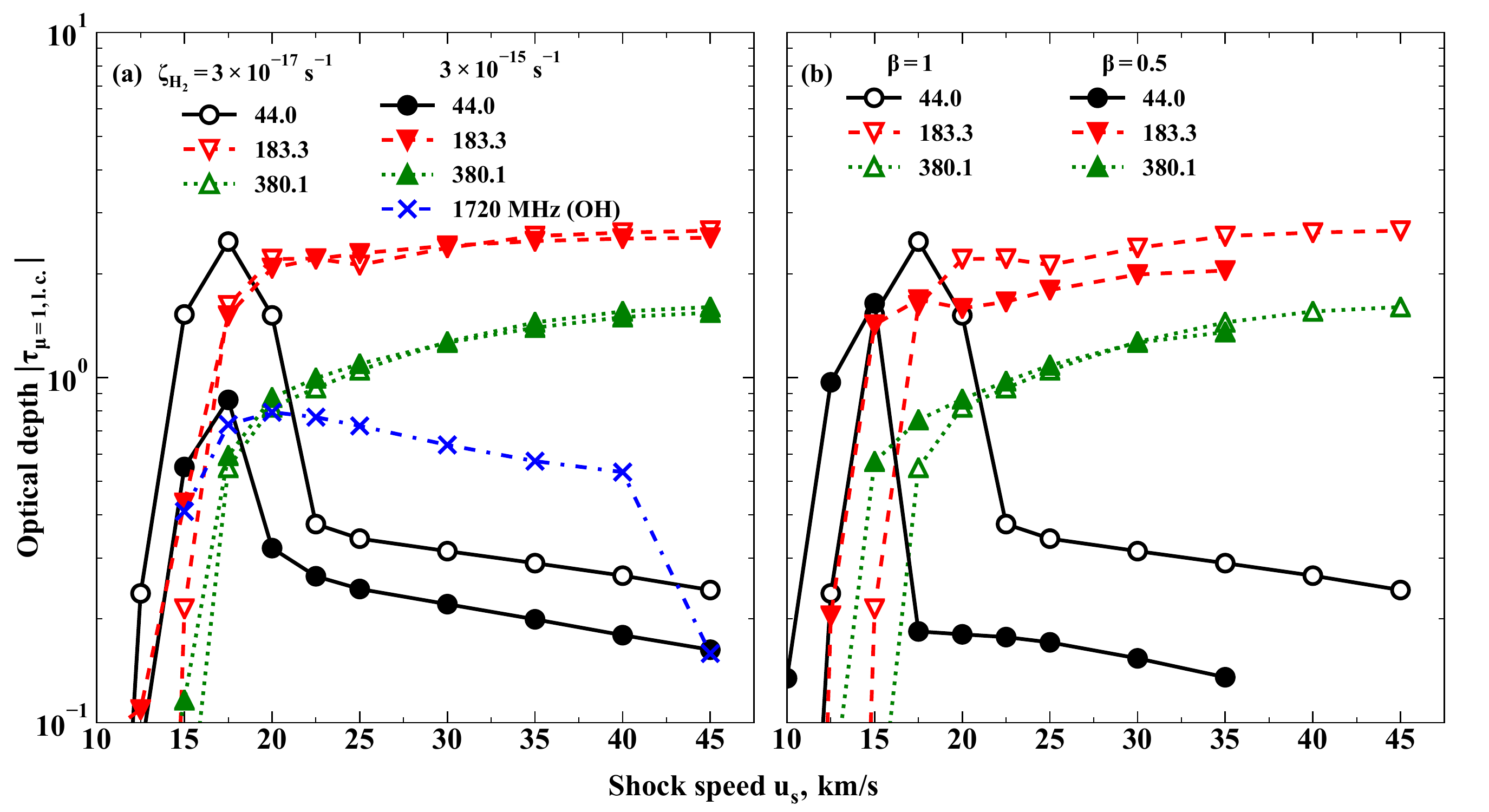}
\caption{Optical depth $\vert \tau_{\rm \mu=1, l.c.}\vert$ for the H$_2$O (183.3 and 380.1~GHz), CH$_3$OH (44.0~GHz), and OH (1720~MHz) maser transitions versus shock speed. The preshock gas density is $n_{\rm H_2,0} = 10^5$~cm$^{-3}$. Panel (a) presents the results of our calculations for two values of the cosmic-ray ionization rate, $\zeta_{\rm H_2} = 3 \times 10^{-17}$ and $3 \times 10^{-15}$~s$^{-1}$ ($\beta = 1$). Panel (b) presents the results of our calculations for two values of the magnetic field, $\beta = 0.5$ and 1 ($\zeta_{\rm H_2} = 3 \times 10^{-17}$~s$^{-1}$).}
\end{figure}

In Fig.~5 the optical depth $\vert \tau_{\rm \mu=1, l.c.}\vert$ is plotted against the shock speed for the H$_2$O, CH$_3$OH, and OH maser transitions; the preshock gas density is $n_{\rm H_2,0} = 10^5$~cm$^{-3}$. Figure~5a presents the results of our calculations for two values of the cosmic-ray ionization rate: $\zeta_{\rm H_2} = 3 \times 10^{-17}$ and $3 \times 10^{-15}$~s$^{-1}$. The results of our calculations for the OH transition between the sublevels of the ground rotational state $^2\Pi_{3/2}$~$j = 3/2$ at 1720 MHz were taken from Nesterenok (2022). At high cosmic-ray ionization rates methanol is destroyed in the postshock region in ion--molecule reactions and reactions of photodissociation by cosmic-ray-induced ultraviolet radiation. The optical depths in the CH$_3$OH maser transitions are considerably smaller at high values of the gas ionization rate than those at low ones (Fig.~5a) (see also Nesterenok (2022)). In contrast, the OH molecule is formed in H$_2$O photodissociation reactions and ion--molecule reactions involving H$_3$O$^+$. High gas ionization rates, $\zeta_{\rm H_2} = 10^{-15}$~s$^{-1}$, are needed for the existence of OH 1720~MHz maser emission. The fraction of the H$_2$O molecules destroyed in the maser zone as a result of these reactions is $\sim 5$\% (for $\zeta_{\rm H_2} = 3 \times 10^{-15}$~s$^{-1}$). Therefore, the influence of the gas ionization rate on the optical depth for the H$_2$O maser transitions is minor. Figure~5b presents the optical depths in the H$_2$O and CH$_3$OH maser lines for two values of the magnetic field ($\beta = 0.5$ and 1), while the gas ionization rate in both cases is $\zeta_{\rm H_2} = 3 \times 10^{-17}$ s$^{-1}$. The lower the value of the magnetic field, the narrower the shock front and the higher gas temperature at the shock front. As a result, the optical depths in the H$_2$O and CH$_3$OH lines are smaller in the case of a weaker magnetic field. The icy mantles of dust grains are sputtered at lower shock speeds. Therefore, the curve of the dependence of the optical depth on the shock speed is shifted leftward for $\beta = 0.5$ (Fig.~5b).

Table~3 gives the brightness temperature of masers calculated from Eq.~(\ref{eq_br_temp}) for the shock parameters $n_{\rm H_2,0} = 10^5$~cm$^{-3}$, $u_{\rm s} = 17.5$~km~s$^{-1}$, $\beta = 1$, and two values of the gas ionization rate, $\zeta_{\rm H_2} = 3 \times 10^{-17}$ and $3 \times 10^{-15}$ s$^{-1}$. The parameter $a = 1/\mu$ was chosen to be 5 in these estimates (where $\mu$ is the cosine of the angle between the line of sight and the shock velocity direction). The background radiation temperature was set equal to $T_{\rm bg} = 3$~K for CH$_3$OH and H$_2$O masers and $T_{\rm bg} = 50$~K for an OH 1720~MHz maser (Hoffman et al. 2005). According to our estimates, the CH$_3$OH masers pass to the saturation regime when the brightness temperature becomes $T_{\rm b,sat} \sim 10^7$~K, while the para-H$_2$O 183.3~GHz maser becomes saturated at $T_{\rm b,sat} \sim 10^9$~K. The lower limits on the brightness temperature given in Table~3 are equal to the maximum brightness temperature of a maser in the unsaturated regime $T_{\rm b,sat}$. In this case, the maser saturation should be taken into account in the brightness temperature calculations, which is beyond the scope of this study. Thus, for the gas ionization rate $\zeta_{\rm H_2} = 3 \times 10^{-15}$~s$^{-1}$, the preshock gas density $n_{\rm H_2,0} = 10^5$~cm$^{-3}$, and the shock speed $u_{\rm s} \approx 20$~km~s$^{-1}$ a coexistence of class I CH$_3$OH, H$_2$O (183.3~GHz), and OH (1720~MHz) masers in one source is possible (provided that the shock velocity direction is perpendicular to the line of sight, $a \sim 5$).

~\\
{{\bf Table 3.} Brightness temperature of OH, CH$_3$OH, and H$_2$O masers} \\ [5pt]
\label{table3}
\vspace{5mm}
\begin{tabular}{lcc}
\hline
Transition & $\zeta_{\rm H_2} = 3 \times 10^{-17}$~s$^{-1}$ & $\zeta_{\rm H_2} = 3 \times 10^{-15}$~s$^{-1}$ \\ [3pt]
\hline
1720~MHz (OH) & -- & $5 \times 10^4$~K \\ [3pt]
44.0~GHz (CH$_3$OH) &  $> 10^7$~K & $\sim 10^7$~K \\ [3pt]
36.1~GHz (CH$_3$OH) &  $> 10^7$~K & $10^4$~K \\ [3pt]
183.3~GHz (H$_2$O) & $2 \times 10^8$~K & $5 \times 10^8$~K \\ [3pt]
\hline
\end{tabular}
~\\
~\\
The shock parameters are $n_{\rm H_2,0} = 10^5$~cm$^{-3}$, $u_{\rm s} = 17.5$~km~s$^{-1}$, and $\beta = 1$; the ratio of the maser amplification path length to the shock width is $1/\mu = 5$. The lower limit on the brightness temperature for CH$_3$OH masers implies that the masers are saturated.
~\\

\section{Discussion}
\subsection{Comparison with the Results of Previous Studies}
Kaufman and Neufeld (1996a) published a C-type shock model and studied the generation of H$_2$O maser emission in shocks of this type. They considered preshock gas densities $n_{\rm H_2,0} = 10^7-10^{9.5}$~cm$^{-3}$, but ignored the interaction of dust grains and gas (adsorption, desorption, the sputtering of icy grain mantles). The gas-phase O-to-H$_2$O conversion reactions were the source of H$_2$O in the gas in their model, whereas in our model H$_2$O is formed on dust grains. At the shock front H$_2$O ends up in the gas as a result of the sputtering of icy grain mantles; the gas-phase H$_2$O formation reactions also make a contribution. Figure 6 shows the average Sobolev optical depth $\bar{\tau}_{\rm S}$ for the H$_2$O 183.3 and 380.1~GHz maser transitions obtained in our calculations and in Kaufman and Neufeld (1996a) (see Fig.~8 in their paper). The results are presented for the preshock gas density $n_{\rm H_2,0} = 10^7$~cm$^{-3}$ (for the determination of the Sobolev optical depth and the method of averaging this parameter in the maser zone, see Kaufman and Neufeld 1996a and 1996b). The relative H$_2$O abundance in the postshock gas is $x_{\rm H_2O} \approx 4 \times 10^{-4}$ in the model of Kaufman and Neufeld (1996a, 1996b). In our model the maximum relative H$_2$O abundance in the cooling postshock gas is $x_{\rm H_2O} \approx 10^{-4}$ -- a factor of 4 lower. In addition, according to our calculations, the width of the postshock region where the maser emission is generated is a factor of $1.5-5$ smaller than that in the model of Kaufman and Neufeld (1996a, 1996b). The lowest shock speed at which the icy mantles of dust grains are sputtered is $17.5-20$~km~s$^{-1}$ for $n_{\rm H_2,0} = 10^7$~cm$^{-3}$ -- this explains the absence of maser emission at low shock speeds in our model. At low shock speeds no evaporation of the icy mantles of dust grains due to grain heating occurs, since the dust temperature is not high enough in our shock model ($T_{\rm d} \lesssim 30$~K for $u_{\rm s} = 10$~km~s$^{-1}$). However, this effect can take place for other dust model parameters or at a higher preshock gas density (Hartquist et al. 1995). At shock speeds $u_{\rm s} > 30$~km~s$^{-1}$ the dissociation of H$_2$ molecules occurs at the shock front, and the shock becomes a J-type shock. These effects explain the difference between the results of our calculations and the results of Kaufman and Neufeld (1996a). Flower and Pineau des For\^{e}ts (2010) also studied the H$_2$O excitation and emission in C-type shocks, but the results for inverted transitions were not discussed in their paper.

Cernicharo et al. (1994) performed numerical simulations of the pumping of para-H$_2$O masers using the large velocity gradient method to solve the radiative transfer equation. They showed that there is level population inversion for the para-H$_2$O 183.3 and 325.1~GHz transitions at relatively low temperatures and gas densities: $T_{\rm g} \approx 100$~K and $n_{\rm H_2} \gtrsim 10^5$~cm$^{-3}$. It also follows from our calculations that efficient pumping of para-H$_2$O 183.3~GHz and ortho-H$_2$O 380.1 and 448.0~GHz masers occurs at low gas densities: $\vert \tau_{\rm \mu=1, l.c.}\vert \sim 1$ for the preshock gas density $n_{\rm H_2,0} = 10^5$~cm$^{-3}$ (Fig.~2). There is energy level population inversion for the para-H$_2$O 325.1~GHz transition for $n_{\rm H_2,0} = 10^5$~cm$^{-3}$, but the optical depth is small, $\vert \tau_{\rm \mu=1, l.c.}\vert < 0.1$. High gas densities, $n_{\rm H_2,0} \gtrsim 10^6$~cm$^{-3}$ (Figs.~2 and 3), are needed for the generation of an intense ortho-H$_2$O 22.23~GHz maser emission ($\vert \tau_{\rm \mu=1, l.c.}\vert \sim 1$).

\begin{figure}[h]
\centering
\includegraphics[width = 0.6\textwidth]{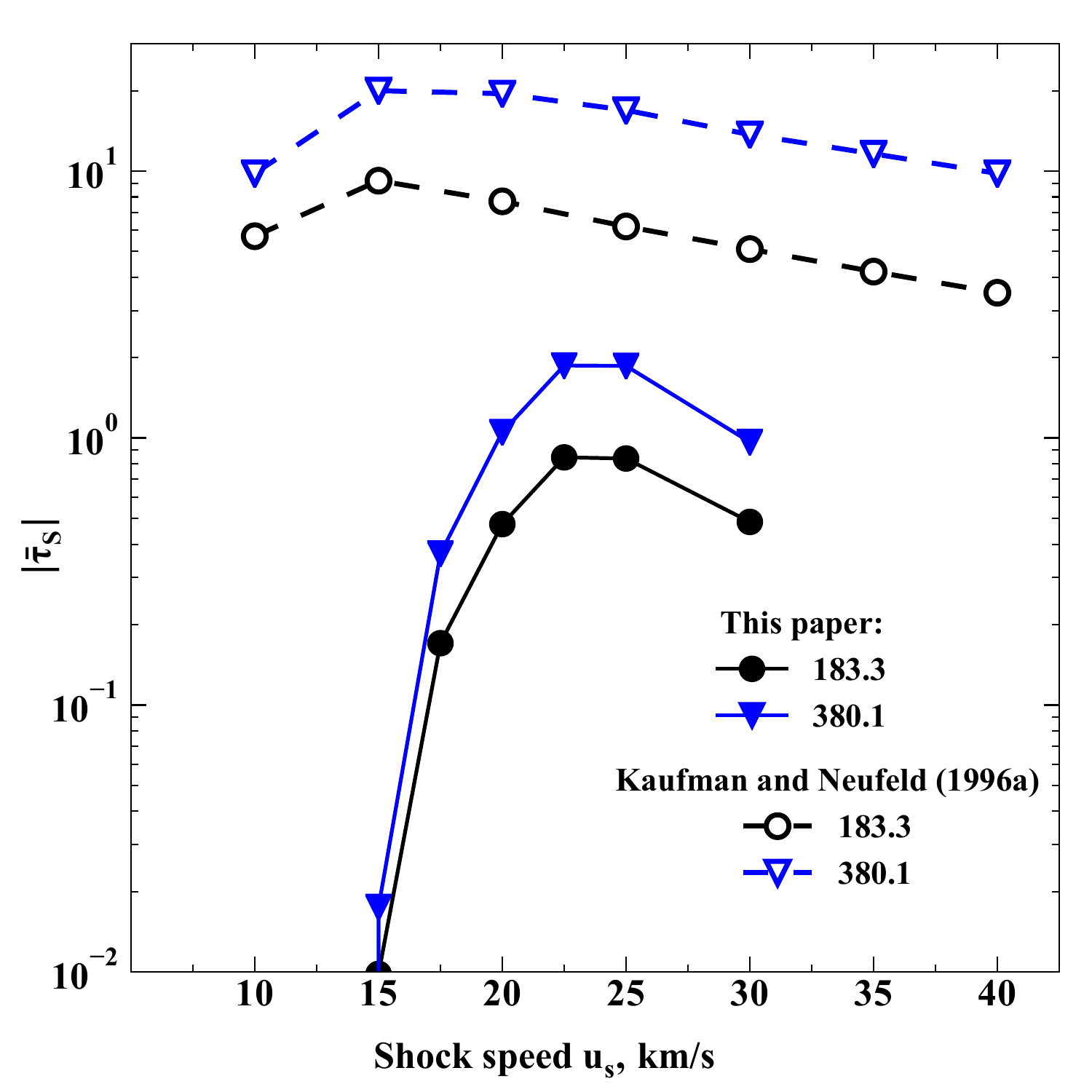}
\caption{Average Sobolev optical depth $\bar{\tau}_{\rm S}$ for the H$_2$O 183.3 and 380.1~GHz maser transitions versus shock speed. The preshock gas density is $n_{\rm H_2,0} = 10^7$~cm$^{-3}$. The results of our calculations and the results from Kaufman and Neufeld (1996a) are shown.}
\end{figure}

\subsection{Observations of H$_2$O 183.3, 380.1, and 448.0~GHz Masers}

Cernicharo et al. (1990, 1994) discovered a spatially distributed para-H$_2$O 183.3~GHz maser emission in Orion A-IRc2 using observations with the IRAM telescope. The size of the region from where the emission comes is $80''$ or $\approx 0.2$~pc, which is much larger than the sizes of the maser spots observed in the 22.23~GHz line in the same object, $\sim 10^{14}-10^{15}$~cm (Genzel et al. 1981). Doty (2000) used the model of a molecular core with a protostar at the center to study the excitation of the para-H$_2$O 183.3~GHz maser line. Doty (2000) showed that a high relative H$_2$O abundance, $x_{\rm H_2O} \sim 10^{-5}$, much higher than the relative H$_2$O abundance in the gas phase in the cold parts of molecular cores, $x_{\rm H_2O} \lesssim 10^{-7}$ (van Dishoeck et al. 2013), is needed to explain the observational data in Orion A-IRc2. This implies that the mechanisms of H$_2$O liberation from the icy mantles of dust grains, such as shocks, need to be invoked.

The observations of the para-H$_2$O 183.3~GHz emission were carried out toward the low-mass protostars HH7-11, L1448-mm, and Serpens SMM1 (Cernicharo et al. 1996; van Kempen et al. 2009). Cernicharo et al. (1996) published the observations of the para-H$_2$O 183.3~GHz emission toward the group of Herbig-Haro objects HH7-11 performed with the IRAM telescope. The emission is generated in a vast region ($>10''$ or $>0.015$~pc) that spatially coincides with the flows near the close binary system SVS~13 visible in CO lines. The emission spectrum has a high-velocity component; the spatial emission peak of this component coincides with HH11. The variability of the radiation intensity points to the maser effect. The brightness temperature of the emission component is $\sim 10$~K. If the emission is assumed to be generated in compact sources whose sizes are much smaller than the telescope's angular resolution ($15''$ or 0.02~pc), then the brightness temperature of the 183.3~GHz masers is $T_{\rm b} >> 10$~K. The H$_2$O masers at 22.23~GHz are located near SVS~13 within $0.3''$ (Rodr{\'{\i}}guez et al. 2002).

Van Kempen et al. (2009) published the observations in the 183.3~GHz line toward the low-mass protostar Serpens SMM1 performed with SMA. The SMA beam sizes were $3'' \times 4''$, corresponding to a linear distance in the source $\approx 1500$~AU (if the distance to the object is taken to be 440~pc; Ortiz-Le{\'o}n et al. 2017). Three spatial emission components that coincide with the flow from the protostar and are at a distance of $1500-3500$~AU from it were detected in Serpens SMM1. The brightness temperature of the components lies in the range $1000-2000$~K, where it was assumed in the estimates that the emission fills completely the SMA beam. If the sizes of the 183.3~GHz emission region is assumed to be $10^{15}$~cm (corresponding to the shock model with the preshock gas density $n_{\rm H_2,0} = 10^5$~cm$^{-3}$), then the brightness temperature in the maser line for the brightest component is $T_b \approx 10^6$~K. Such brightness temperatures of the H$_2$O 183.3~GHz maser are reproduced in a C-type shock model with parameters $n_{\rm H_2,0} = 10^5$~cm$^{-3}$, $u_{\rm s} \gtrsim 17.5$~km~s$^{-1}$, and $a \approx 3$. Moscadelli et al. (2006) observed H$_2$O 22.23~GHz masers toward Serpens SMM1 with VLBA. The 22.23~GHz maser emission sources are located at a distance $\approx 10-20$~AU from the protostar (probably, inside an accretion disk), while the sizes of the maser spots are $<5$~AU. Thus, the para-H$_2$O 183.3~GHz masers, along with the class I CH$_3$OH masers, are indicators of the gas flows interacting with the protostar envelope and the interstellar medium, whereas the H$_2$O 22.23~GHz masers emerge in the immediate vicinity of protostars.

The generation of para-H$_2$O 183.3~GHz and ortho-H$_2$O 380.1 and 448.0~GHz maser emission is possible at lower gas densities than those for the ortho-H$_2$O 22.23~GHz transition. The emission in these lines provides an additional possibility for diagnosing the physical conditions in astrophysical objects (for example, K\"{o}nig et al. 2017). The ortho-H$_2$O 380.1 and 448.0~GHz transitions cannot be observed with ground-based telescopes in astrophysical objects of our Galaxy due to absorption in the Earth's atmosphere. However, these transitions can be observed toward galaxies in the local Universe and at cosmological distances, where the emission in these lines is shifted to a frequency range accessible to observation (Pereira-Santaella et al. 2017; Kuo et al. 2019; Yang et al. 2020). In particular, using the ALMA radio interferometer, Kuo et al. (2019) observed ortho-H$_2$O 380.1~GHz emission toward the lensed quasar QSO MG J0414+0534 at redshift $z = 2.639$. The recorded emission may be a maser one in nature -- according to the estimates by Kuo et al. (2019), the isotropic (unlensed) line luminosity is $ \approx 5 \times 10^6 L_{\odot}$.

\subsection{Absence of H$_2$O 22.23~GHz Masers Associated with Supernova Remnants}
Claussen et al. (1999) searched for H$_2$O 22.23~GHz emission toward three supernova remnants in which the OH maser emission at 1720~MHz was known: W28, W44, and IC 443. Woodall and Gray (2007) searched for 22.23~GHz emission toward 18 supernova remnants (they also included supernova remnants where no OH emission was recorded in their sample). The 22.23~GHz emission was not detected in any of the sources. Using the C- and J-type shock models, Woodall and Gray (2007) performed numerical simulations of the pumping of H$_2$O 22.23~GHz masers. The preshock gas density in their numerical simulations varied in the range $n_{\rm H_2,0} = 10^3-10^5$~cm$^{-3}$ -- the collisional pumping of OH masers at 1720~MHz is efficient precisely at these preshock gas densities. Woodall and Gray (2007) showed that there is no H$_2$O 22.23~GHz maser emission at these gas densities. The same conclusions follow from our calculations -- the optical depth in the 22.23~GHz line is $\vert \tau_{\rm \mu=1, l.c.}\vert \lesssim 0.05$ for the preshock gas density $n_{\rm H_2,0} = 10^5$~cm$^{-3}$. At the same time, the generation of para-H$_2$O 183.3~GHz maser emission is possible for this preshock gas density (just as in the ortho-H$_2$O 380.1 and 448.0~GHz lines, but these lines is difficult to observe in Galactic objects due to absorption in the Earth's atmosphere). Note that OH 1720~MHz and CH$_3$OH 36.1 and 44.0~GHz maser emission was observed near the supernova remnants W28 and W44 (Pihlstr\"{o}m et al. 2014; McEwen et al. 2016).

\section{Conclusions}
We investigated the collisional pumping of H$_2$O and CH$_3$OH masers in C-type shocks. Within the models considered here we showed that the para-H$_2$O 183.3~GHz and ortho-H$_2$O 380.1 and 448.0~GHz transitions could be inverted at relatively low preshock gas densities, $n_{\rm H_2,0} \approx 10^5$~cm$^{-3}$. The generation of H$_2$O and CH$_3$OH maser emission in the same postshock region is possible at these gas densities and shock velocities $u_{\rm s} = 17.5-22.5$~km~s$^{-1}$. We showed that the effect of the ortho-/para-H$_2$ ratio on the pumping of H$_2$O masers in a shock is minor and it is significant on the pumping of CH$_3$OH masers. No H$_2$O 22.23~GHz maser emission associated with supernova remnants has been detected previously. The relatively low gas densities in shocks in supernova remnants are most likely responsible for the absence of 22.23~GHz maser emission. According to our calculations, for preshock gas densities $n_{\rm H_2,0} \leq 10^5$~cm$^{-3}$ the optical depth in the 22.23~GHz line along the gas flow in the shock is small, $< 0.05$. Our numerical simulations suggest that para-H$_2$O 183.3~GHz emission can be detected in those supernova remnant regions where the 1720~MHz OH and class I CH$_3$OH maser emission is generated. The para-H$_2$O 183.3~GHz maser emission provides an additional possibility to investigate the physical conditions in protostellar flows in star-forming regions and near supernova remnants.

For me it is a great honor to devote this paper to the memory of my teacher and scientific adviser, academician Dmitrii Aleksandrovic Varshalovich (1934--2020) of the Russian Academy of Sciences. Under his leadership I began to investigate the interstellar medium and cosmic masers. Dmitrii Aleksandrovic will always remain in memory as an outstanding scientist and a remarkable man.

\section{References}
\noindent
1. A. C. A. Boogert, P. A. Gerakines, and D. C. B. Whittet,
Ann. Rev. Astron. Astrophys. {\bf 53}, 541 (2015).

\noindent
2. J. Cernicharo, C. Thum, H. Hein, D. John, P. Garcia, and F. Mattioco, Astron. Astrophys. {\bf 231}, L15 (1990).

\noindent
3. J. Cernicharo, E. Gonz\'{a}lez-Alfonso, J. Alcolea, R. Bachiller, and D. John, Astrophys. J. {\bf 432}, L59 (1994).

\noindent
4. J. Cernicharo, R. Bachiller, and E. Gonz\'{a}lez-Alfonso, Astron. Astrophys. {\bf 305}, L5 (1996).

\noindent
5. J. Cernicharo, J. R. Pardo, E. Gonz\'{a}lez-Alfonso, E. Serabyn, T. G. Phillips, D. J. Benford, and D. Mehringer, Astrophys. J. {\bf 520}, L131 (1999).

\noindent
6. M. J. Claussen, W. M. Goss, and D. A. Frail, Astron. J. {\bf 117}, 1387 (1999).

\noindent
7. R. M. Crutcher, B. Wandelt, C. Heiles, E. Falgarone, and T. H. Troland, Astrophys. J. {\bf 725}, 466 (2010).

\noindent
8. A. Dalgarno, Proc. Natl. Acad. Sci. U. S. A. {\bf 103}, 12269 (2006).

\noindent
9. F. Daniel and J. Cernicharo, Astron. Astrophys. {\bf 553}, A70 (2013).

\noindent
10. F. Daniel, A. Faure, P. J. Dagdigian, M.-L. Dubernet, F. Lique, and G. Pineau des For\^{e}ts, Mon. Not. R. Astron. Soc. {\bf 446}, 2312 (2015).

\noindent
11. E. F. van Dishoeck, E. Herbst, and D. A. Neufeld, Chem. Rev. {\bf 113}, 9043 (2013).

\noindent
12. S. D. Doty, Astrophys. J. {\bf 535}, 907 (2000).

\noindent
13. B. T. Draine and C. F. McKee, Ann. Rev. Astron. Astrophys. {\bf 31}, 373 (1993).

\noindent
14. A. E. Dudorov, Sov. Astron. {\bf 35}, 342 (1991).

\noindent
15. M. Elitzur, D. J. Hollenbach, and C. F. McKee, Astrophys. J. {\bf 346}, 983 (1989).

\noindent
16. M. Emprechtinger, D. C. Lis, R. Rolffs, P. Schilke, R. R. Monje, C. Comito, C. Ceccarelli, D. A. Neufeld, et al., Astrophys. J. {\bf 765}, 61 (2013).

\noindent
17. A. Faure, N. Crimier, C. Ceccarelli, P. Valiron, L. Wiesenfeld, and M. L. Dubernet, Astron. Astrophys. {\bf 472}, 1029 (2007).

\noindent
18. A. Faure and E. Josselin, Astron. Astrophys. {\bf 492}, 257 (2008).

\noindent
19. D. R. Flower and G. Pineau des For\^{e}ts, Mon. Not. R. Astron. Soc. {\bf 406}, 1745 (2010).

\noindent
20. R. Genzel, M. J. Reid, J. M. Moran, and D. Downes, Astrophys. J. {\bf 244}, 884 (1981).

\noindent
21. I. E. Gordon, L. S. Rothman, R. J. Hargreaves, R. Hashemi, E. V. Karlovets, F. M. Skinner, E. K. Conway, C. Hill, et al., J. Quant. Spectrosc. Radiat. Transfer {\bf 277}, 107949 (2022).

\noindent
22. M. D. Gray, A. Baudry, A. M. S. Richards, E. M. L. Humphreys, A. M. Sobolev, and J. A. Yates, Mon. Not. R. Astron. Soc. {\bf 456}, 374 (2016).

\noindent
23. M. D. Gray, S. Etoka, A. M. S. Richards, and B. Pimpanuwat,
Mon. Not. R. Astron. Soc. {\bf 513}, 1354 (2022).

\noindent
24. S. Green, S. Maluendes, and A. D. McLean, Astrophys. J. Suppl. Ser. {\bf 85}, 181 (1993).

\noindent
25. T. W. Hartquist, K. M. Menten, S. Lepp, and A. Dalgarno, Mon. Not. R. Astron. Soc. {\bf 272}, 184 (1995).

\noindent
26. I. M. Hoffman, W. M. Goss, C. L. Brogan, and M. J. Claussen, Astrophys. J. {\bf 627}, 803 (2005).

\noindent
27. D. Hollenbach, M. Elitzur, and C. F. McKee, Astrophys. J. {\bf 773}, 70 (2013).

\noindent
28. D. G. Hummer and G. B. Rybicki, Astrophys. J. {\bf 293}, 258 (1985).

\noindent
29. M. J. Kaufman and D. A. Neufeld, Astrophys. J. {\bf 456}, 250 (1996a).

\noindent
30. M. J. Kaufman and D. A. Neufeld, Astrophys. J. {\bf 456}, 611 (1996b).

\noindent
31. T. A. van Kempen, D. Wilner, and M. Gurwell, Astrophys. J. {\bf 706}, L22 (2009).

\noindent
32. S. K\"{o}nig, S. Mart{\'{\i}}n, S. Muller, J. Cernicharo, K. Sakamoto, L. K. Zschaechner, E. M. L. Humphreys, T. Mroczkowski, et al., Astron. Astrophys. {\bf 602}, A42 (2017).

\noindent
33. C.-Y. Kuo, S. H. Suyu, V. Impellizzeri, and J. A. Braatz, Publ. Astron. Soc. Jpn. {\bf 71}, 57 (2019).

\noindent
34. S. Leurini, K. M. Menten, and C. M. Walmsley, Astron. Astrophys. {\bf 592}, A31 (2016).

\noindent
35. B. C. McEwen, Y. M. Pihlstr\"{o}m, and L. O. Sjouwerman, Astrophys. J. {\bf 793}, 133 (2014).

\noindent
36. B. C. McEwen, Y. M. Pihlstr\"{o}m, and L. O. Sjouwerman, Astrophys. J. {\bf 826}, 189 (2016).

\noindent
37. L. Moscadelli, L. Testi, R. S. Furuya, C. Goddi, M. Claussen, Y. Kitamura, and A. Wootten, Astron. Astrophys. {\bf 446}, 985 (2006).

\noindent
38. A. V. Nesterenok, Astron. Lett. {\bf 39}, 717 (2013).

\noindent
39. A. V. Nesterenok, Mon. Not. R. Astron. Soc. {\bf 455}, 3978 (2016).

\noindent
40. A. V. Nesterenok, Astrophys. Space Sci. {\bf 363}, 151 (2018).

\noindent
41. A. V. Nesterenok, Astron. Lett. {\bf 46}, 449 (2020).

\noindent
42. A. V. Nesterenok, J. Phys.: Conf. Ser. {\bf 2103}, 012012 (2021).

\noindent
43. A. V. Nesterenok, Mon. Not. R. Astron. Soc. {\bf 509}, 4555 (2022).

\noindent
44. A. V. Nesterenok and D. A. Varshalovich, Astron. Lett. {\bf 37}, 456 (2011).

\noindent
45. A. V. Nesterenok, D. Bossion, Y. Scribano, and F. Lique, Mon. Not. R. Astron. Soc. {\bf 489}, 4520 (2019).

\noindent
46. D. A. Neufeld and G. J. Melnick, Astrophys. J. {\bf 368}, 215 (1991).

\noindent
47. D. A. Neufeld, G. J. Melnick, M. J. Kaufman, H. Wiesemeyer, R. G\"{u}sten, A. Kraus, K. M. Menten, O. Ricken, et al., Astrophys. J. {\bf 843}, 94 (2017).

\noindent
48. G. N. Ortiz-Le\'{o}n, S. A. Dzib, M. A. Kounkel, L. Loinard, A. J. Mioduszewski, L. F. Rodr{\'{\i}}guez, R. M. Torres, G. Pech, et al., Astrophys. J. {\bf 834}, 143 (2017).

\noindent
49. M. Pereira-Santaella, E. Gonz{\'a}lez-Alfonso, A. Usero, S. Garc{\'{\i}}a-Burillo, J. Mart{\'{\i}}n-Pintado, L. Colina, A. Alonso-Herrero, S. Arribas, et al., Astron. Astrophys. {\bf 601}, L3 (2017).

\noindent
50. Y. M. Pihlstr\"{o}m, L. O. Sjouwerman, D. A. Frail, M. J. Claussen, R. A. Mesler, and B. C. McEwen, Astron. J. {\bf 147}, 73 (2014).

\noindent
51. L. F. Rodr{\'{\i}}guez, G. Anglada, J. M. Torrelles, J. E. Mendoza-Torres, A. D. Haschick, and P. T. P. Ho, Astron. Astrophys. {\bf 389}, 572 (2002).

\noindent
52. S. V. Salii, A. M. Sobolev, and N. D. Kalinina, Astron. Rep. {\bf 46}, 955 (2002).

\noindent
53. C. N. Shingledecker, J. B. Bergner, R. Le Gal, K. I. \"{O}berg, U. Hincelin, and E. Herbst, Astrophys. J. {\bf 830}, 151 (2016).

\noindent
54. V. S. Strelnitskii, Sov. Phys. Usp. {\bf 17}, 507 (1975).

\noindent
55. D. A. Varshalovich, A. V. Ivanchik, and N. S. Babkovskaya, Astron. Lett. {\bf 32}, 29 (2006).

\noindent
56. M. A. Voronkov, J. L. Caswell, S. P. Ellingsen, J. A. Green, and S. L. Breen, Mon. Not. R. Astron. Soc. {\bf 439}, 2584 (2014).

\noindent
57. N. Watanabe and A. Kouchi, Astrophys. J. {\bf 571}, L173 (2002).

\noindent
58. J. M. Woodall and M. D. Gray, Mon. Not. R. Astron. Soc. {\bf 378}, L20 (2007).

\noindent
59. C. Yang, E. Gonz{\'a}lez-Alfonso, A. Omont, M. Pereira-Santaella, J. Fischer, A. Beelen, and R. Gavazzi, Astron. Astrophys. {\bf 634}, L3 (2020).

\noindent
60. J. A. Yates, D. Field, and M. D. Gray, Mon. Not. R. Astron. Soc. {\bf 285}, 303 (1997).

~\\
~\\
Translated by V. Astakhov

\end{document}